\documentclass[%
 reprint,
 showpacs,preprintnumbers,
 nofootinbib,
 amsmath,amssymb,
 aps,
 prc,
 floatfix,
]{revtex4-1}
\usepackage{graphicx}

\usepackage{amssymb,amsmath,amstext,amsthm,amsfonts}
\usepackage[utf8]{inputenc}
\usepackage{subcaption}
\usepackage{float}
\usepackage{url}
\usepackage{todonotes}

\usepackage{changepage}                 
\usepackage[export]{adjustbox}[2011/08/13]

\newlength{\offsetpage}
\newenvironment{widepage}{\begin{adjustwidth}{-\offsetpage}{-\offsetpage}%
    \addtolength{\textwidth}{2\offsetpage}}%
{\end{adjustwidth}}

\begin{document}
	\title{Sensitivity of Neutrino-Nucleus Interaction Measurements to 2p2h Excitations}
	
	\author{S. Dolan}
		\email[Contact e-mail: ]{Stephen.Dolan@llr.in2p3.fr}
	\affiliation{Ecole Polytechnique, IN2P3-CNRS, Laboratoire Leprince-Ringuet, Palaiseau 91120, France }
	\affiliation{IRFU, CEA Saclay, Gif-sur-Yvette 91190, France}
 
	\author{U. Mosel}
	\email[Contact e-mail: ]{mosel@physik.uni-giessen.de}
	\affiliation{Institut f\"ur Theoretische Physik, Universit\"at Giessen, Giessen D-35392, Germany}
	
	\author{K. Gallmeister}
	\affiliation{Institut f\"ur Theoretische Physik, Johann Wolfgang Goethe-Universit\"at Frankfurt, Frankfurt am Main D-60438, Germany}
	
	\author{L. Pickering}
	\affiliation{Michigan State University, Department of Physics and Astronomy,  East Lansing 48823, Michigan, U.S.A.}
	
	\author{S. Bolognesi}
	\affiliation{IRFU, CEA Saclay, Gif-sur-Yvette 91190, France}

\begin{abstract}
		We calculate the charged-current cross sections obtained at the T2K off-axis near detector for $\nu_\mu$-induced events without pions and any number of protons in the final state using transport theory as encoded in the  GiBUU model. In a comparison with recent T2K data the strength of the 2p2h multinucleon correlations is determined. Linking this to the isospin ($\mathcal{T}$) of the initial nuclear state, it is found that $\mathcal{T}=0$ leads to a significantly better fit of the recent cross sections obtained by T2K, thus achieving consistency of the 2p2h multi-nucleon correlation contributions between electron-nucleus and neutrino-nucleus reactions.
\end{abstract}

\maketitle
\section{Introduction}	
The accurate characterisation of (sub)GeV-scale charged-current neutrino-nucleus interactions through differential cross-section measurements is essential for long-baseline neutrino oscillation experiments to determine the neutrino mixing parameters. The inclusive cross sections are comprised of a sum of quasielastic (CCQE) scattering, 2p2h multi-nucleon interactions and pion production processes. The 2p2h component is particularly interesting as theoretical models differ substantially in their predicted 2p2h-strengths and the systematic uncertainty applied to cover this in oscillation analyses can be one of the largest~\cite{Abe:2017vif, Duffy:2016phs}. Moreover, the experimental picture of 2p2h interactions obtained from MiniBooNE, MINERvA and T2K measurements is confused. Results from MiniBooNE showed an excess over predictions in a CCQE cross-section measurement but are consistent with either the inclusion of 2p2h or with an enhanced CCQE contribution~\cite{AguilarArevalo:2010zc, Nieves:2011yp}. Following this, results from MINERvA measuring muon and proton final states found no evidence of a 2p2h contribution~\cite{Walton:2014esl}, in contrast with their results measuring muon-only final states which suggest a need for 2p2h when considering a relativistic Fermi-gas model for the CCQE component~\cite{Fiorentini:2013ezn}. However, comparing this result with more sophisticated nuclear models~\cite{Mosel:2014lja} yields a conclusion consistent with only a CCQE contribution. MINERvA recently updated their flux prediction and a comparison to a similar analysis with this update is then consistent with a 2p2h contribution~\cite{Megias:2016fjk}. More recent MINERvA measurements~\cite{Gran:2018fxa, Patrick:2018gvi, Rodrigues:2015hik} are not consistent with the 2p2h model of the IFIC Valencia group~\cite{Nieves:2011pp} but require an empirical enhancement. Previous results from T2K~\cite{Abe:2016tmq} suggest the need for a 2p2h contribution but experimental uncertainties are too large to offer a firm quantification. In none of these studies was the connection to complementary electron-induced data directly exploited.

To clarify the experimental situation, recent results from T2K~\cite{Abe:2018pwo} attempt to provide a deeper probe of 2p2h and other nuclear-medium effects by analyzing the kinematics of both the outgoing muon and proton(s) together following neutrino interactions on a C$_8$H$_8$ target. In this paper we analyze these results with the intent of characterizing the 2p2h strength. By invoking consistency with electron data we then determine the isospin $\mathcal{T}$ of the initial nuclear configuration.

\section{Model}
We use the non-equilibrium Green's function method encoded in the quantum-kinetic transport code GiBUU. Its theoretical foundations as well as its numerical implementation are described in some detail in \cite{Buss:2011mx}. Since the present paper focuses on 2p2h interactions we repeat here, for easier reference,  some relevant details on the treatment of 2p2h contributions in GiBUU; for further details we refer to~\cite{Gallmeister:2016dnq}.

The neutrino 2p2h cross-section contribution ($\sigma^{2p2h}$) can be written in terms of the neutrino structure functions $W_1^\nu(Q^2,\omega)$ and $W_3^\nu(Q^2,\omega)$ as: 
\begin{eqnarray}  \label{LPnu}
\frac{d^2\sigma^{2p2h}}{d\Omega dE'}
&=& \frac{G^2}{2 \pi^2} E'^2 \cos^2 \frac{\theta}{2} \,\left[2W_1^\nu \left(\frac{Q^2}{2\mathbf{q}^2} + \tan^2\frac{\theta}{2}  \right) \right.  \nonumber \\
& & \mbox{}\left. \mp W_3^\nu \frac{E + E'}{M} \tan^2\frac{\theta}{2}\right] ~.
\end{eqnarray}
Here $G$ is the weak coupling constant; $E'$ and $\theta$ are the outgoing lepton energy and angle respectively; $E$ is the incoming neutrino energy; $Q^2$ is the squared four momentum transfer $Q^2 = \mathbf{q}^2-\omega^2$ and $M$ is the nucleon mass. In Eq.\ (\ref{LPnu}) the 2p2h contribution is purely transverse.

Crucial for the present study is the connection between the electron and the neutrino structure functions derived in Ref.\ \cite{OConnell:1972edu}. There the authors used the Wigner-Eckart theorem to connect the charge-changing transition rates, as they occur in CC neutrino-interactions, with the charge-conserving ones appearing in electron scattering. We exploit this here to link the neutrino 2p2h strength to the much better known electron 2p2h contribution

The 2p2h neutrino structure functions are related to those for electrons, $W_1^e$, by a simple factor that involves kinematical quantities and the coupling constants as well as the isospin $\mathcal{T}$ of the initial state:
\begin{eqnarray}
\label{eq:W1}
W_1^e &=&   G_M^2 \frac{\omega^2}{\mathbf{q}^2}  \, R_T^e \nonumber \\
W_1^\nu &=& \left( G_M^2 \frac{\omega^2}{\mathbf{q}^2} +  G_A^2\right) \, R_T^e \, 2(\mathcal{T} + 1) ~.
\end{eqnarray}
Here $R_T^e$ is a reduced electromagnetic transverse response function and  $G_A(Q^2)$ and $G_M(Q^2)$ are the axial and magnetic form factors, respectively. A similar structure shows up in the V-A interference structure function:
\begin{equation}
\label{eq:W3}
W_3^\nu = 2 G_A G_M  \, R_T^e \, 2(\mathcal{T} + 1) ~.
\end{equation} 
We note that this form (except for the isospin factor)  has also been used in all calculations by the Lyon group \cite{Martini:2009uj}.

The structure of Eq.\ (\ref{eq:W1}) can be understood by noting that $R_T^e$ is a ``reduced electromagnetic structure function'', from which the electromagnetic form factor and a kinematical factor ${\omega}^2/\mathbf{q}^2$ have been removed. The full structure function for neutrinos is then obtained by multiplying $R_T^e$ with the $VV$ and the $AA$ coupling constants. The $VA$ interference shows up in the presence of the product $G_A G_M$ in Eq.\ (\ref{eq:W3}).

The electron structure function $W_1^e$ encoded in GiBUU is obtained from a fit to the dip region in electron scattering cross sections for a wide kinematical range \cite{Bosted:2012qc,Christy:2015}. It is essential to realize that while this contribution is abbreviated by 'MEC' in Ref.\ \cite{Bosted:2012qc} this empirical structure function contains the combined effects of meson exchange currents, of two-nucleon correlations, and of short-range correlations. In this fit the 2p2h contribution was assumed to be purely transverse, in line with Eq.\ (\ref{LPnu}). In Ref.\ \cite{Gallmeister:2016dnq} it was shown that GiBUU describes electron scattering inclusive cross sections on $^{12}C$ with this electron structure function quite well. This is also due to the fact
that the model in the 1p1h sector goes beyond the local Fermi gas because it embeds the nucleons in a binding potential. This latter feature takes care of the essential 1p1h contributions embodied in the mean field potential and removes the unphysical $\delta$-function spikes in the hole spectral function that are present in the Fermi gas.

For isospins $\mathcal{T}=0$ and $\mathcal{T}=1$ the neutrino structure functions, and thus also the cross sections for the 2p2h-channel, differ by just a factor of two. Although the isospin of the $^{12}$C groundstate is $\mathcal{T}=0$, results from the MiniBooNE experiment~\cite{AguilarArevalo:2010zc,Aguilar-Arevalo:2013dva} have suggested that $\mathcal{T}=1$ better describes the data \cite{Gallmeister:2016dnq}. However, as already discussed in~\cite{Gallmeister:2016dnq}, the MiniBooNE results offer sensitivity to 2p2h largely only through the absolute normalisation of the measured cross section. The large overall flux normalisation uncertainty weakens any conclusion regarding the 2p2h strength. It is the aim of the present study to investigate whether the T2K results, of which some offer significant sensitivity to 2p2h through the shape of the measured cross section, are sensitive to this factor of two and, if so, which of the two values is favored. 

In the present calculations we use the GiBUU 2017 version as it can be downloaded from~\cite{gibuuurl}. No special tunes or adjustments were made.

\section{Comparison with experiment}

We compare the aforementioned T2K differential cross-section measurements of interactions that leave no mesons in the final state to GiBUU model predictions of the same topology with $\mathcal{T} =0$ and $\mathcal{T} =1$. The different isospin assignments affect only the 2p2h contribution while all the other components of the cross section remain unaltered. To provide a quantitative metric for comparison $\chi^2$ statistics are calculated. It is essential to use the full experimental covariance matrix due to strong off-diagonal contributions stemming from detector resolution effects and the influence of the highly correlated flux uncertainty on the results. Furthermore, it is important to note that these $\chi^2$ statistics can also suffer from Peelle's Pertinent Puzzle and may therefore over-favour model predictions with a lower overall normalization~\cite{ppp,ppp2}. For this reason we will focus primarily on interpreting comparisons in specific regions of kinematic phase-space, rather than based on the $\chi^2$. The comparisons shown within this section have been made within the NUISANCE framework~\cite{Stowell:2016jfr}. 

\begin{figure*}
\includegraphics[width=1.1\textwidth, center]{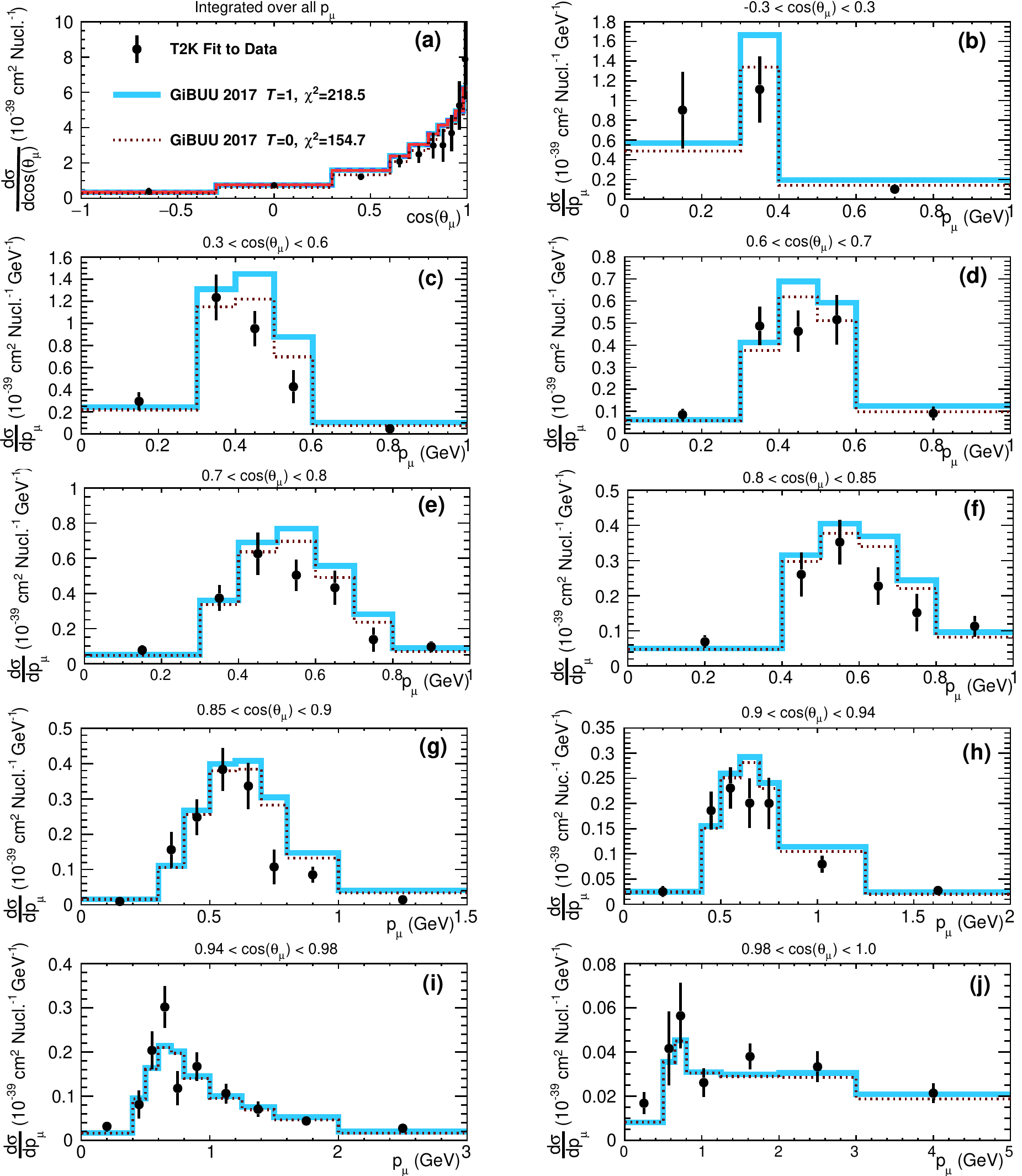}
\caption{The T2K measurement of the double-differential charged-current pionless cross-section when there are no protons with momenta above 500 MeV on a C$_8$H$_8$ target as a function of muon kinematics \cite{Abe:2018pwo} compared with the results of a GiBUU 2017 calculation with $\mathcal{T} =0$ and $\mathcal{T} =1$. Each plot shows the comparison as a function of the outgoing muon momentum within a particular angular slice (the angle is defined with respect to the direction of the incoming neutrino). The $\chi^2$ shown in the legend are calculated for 59 degrees of freedom.}
\label{fig:dd}

\end{figure*}

We start our discussion with the double-differential cross section as a function of the muon kinematics for events with no outgoing protons with momenta above 500 MeV, shown in Fig.~\ref{fig:dd}. 


Here a modest preference for $\mathcal{T}=0$ can be seen, driven by the intermediate momentum bins of the higher angle slices where the 2p2h content is predicted to be largest~\cite{Mosel:2017ssx}. It is interesting to note this prediction of a predominantly high-angle 2p2ph component runs contrary to what it predicted by other 2p2h models, as can be seen in~\cite{Abe:2016tmq} for the Lyon and Valencia models and in~\cite{Megias:2016fjk} for SuSAv2. These other 2p2h models tend to overestimate the forward cross section, which could indicate problems with their longitudinal 2p2h contributions, as has been speculated in Ref.\ \cite{Mosel:2017ssx}.

We also compared the GiBUU prediction to the T2K results with one proton with momentum above 500 MeV, but this shows little sensitivity to variations in $\mathcal{T}$ and so is not shown here. This is partially due to GiBUU predicting a slightly smaller relative contribution of 2p2h interactions but also because the limited experimental statistics only facilitate binning where their contribution is not so concentrated. The smaller 2p2h component is largely because T2K's fairly narrow and low neutrino beam energy (peaked at 0.6 GeV)~\cite{Abe:2012av} allows only a fairly limited kinematic phase-space where a two nucleon excitation can leave one with momentum above 500 MeV.

The T2K analysis went beyond a double-differential cross section measurement by also exploring the correlations between outgoing muon and proton kinematics, which can serve as a more powerful projection of the data when characterizing the 2p2h contribution. One such analysis measures three observables which describe the imbalance between the muon and the highest-momentum proton in the plane transverse to the incoming neutrino.  These ``single-transverse'' observables have previously been demonstrated to act as powerful probes of nuclear-medium effects for measurements of an exclusive interaction mode~\cite{Lu:2015tcr}. They are defined by: 
\begin{eqnarray}
\delta p_T &=& |\mathbf{\delta p}_T| = \left| \mathbf{p}_T^l + \mathbf{p}_T^p \right|     \nonumber \\
\delta \alpha_T &=& {\rm arccos}\left(- \frac{\mathbf{p}_T^l \cdot \mathbf{\delta p}_T}{p_T^l \delta p_T} \right) \ \nonumber \\
\delta \phi_T &=& {\rm arccos}\left( - \frac{\mathbf{p}_T^l \cdot \mathbf{p}_T^p}{p_T^l p_T^p}\right)   ~.
\end{eqnarray}
Here $\mathbf{p}_T^l$ and $\mathbf{p}_T^p$ are the momentum of the outgoing lepton and highest momentum proton in the plane transverse to the incoming neutrino, respectively. 

\begin{figure*} 
\begin{widepage}

\vspace{-0.5cm}
\hspace{0.5cm}
\begin{subfigure}{0.52\textwidth}
\includegraphics[width=\linewidth]{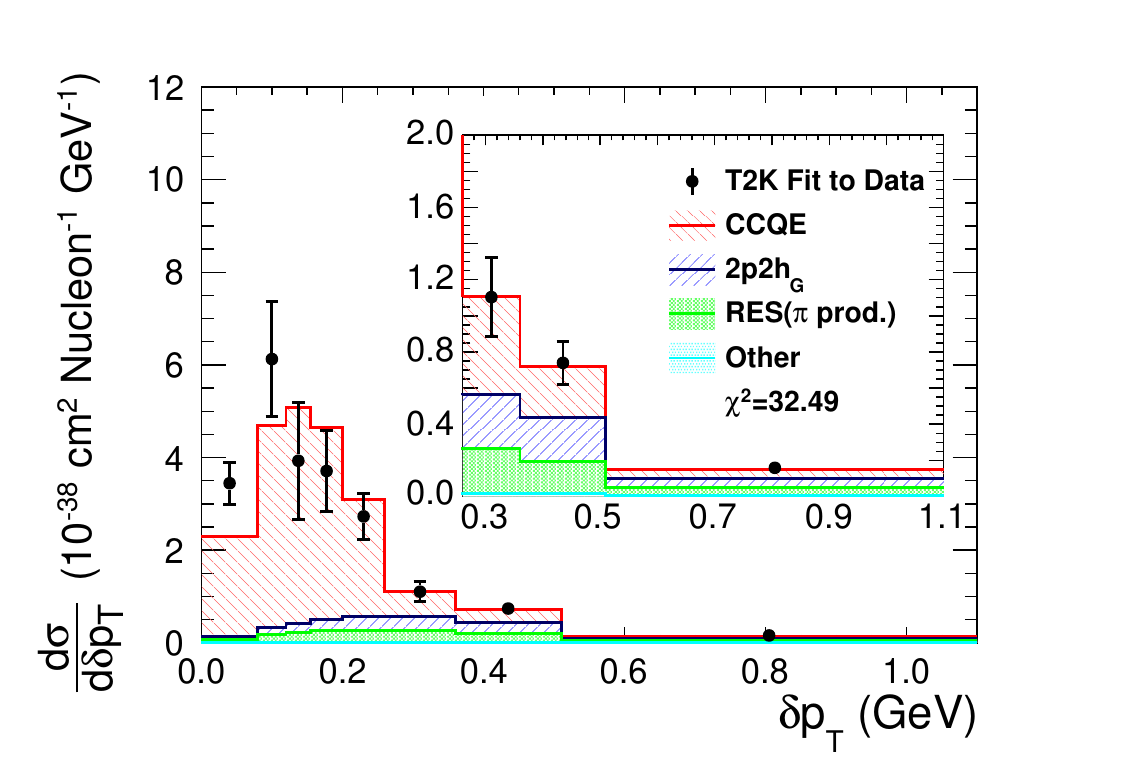}
\caption{$\delta p_{T}$ for $\mathcal{T} =0$} \label{fig:e}
\end{subfigure}\hspace{-1.1cm}
\begin{subfigure}{0.52\textwidth}
\includegraphics[width=\linewidth]{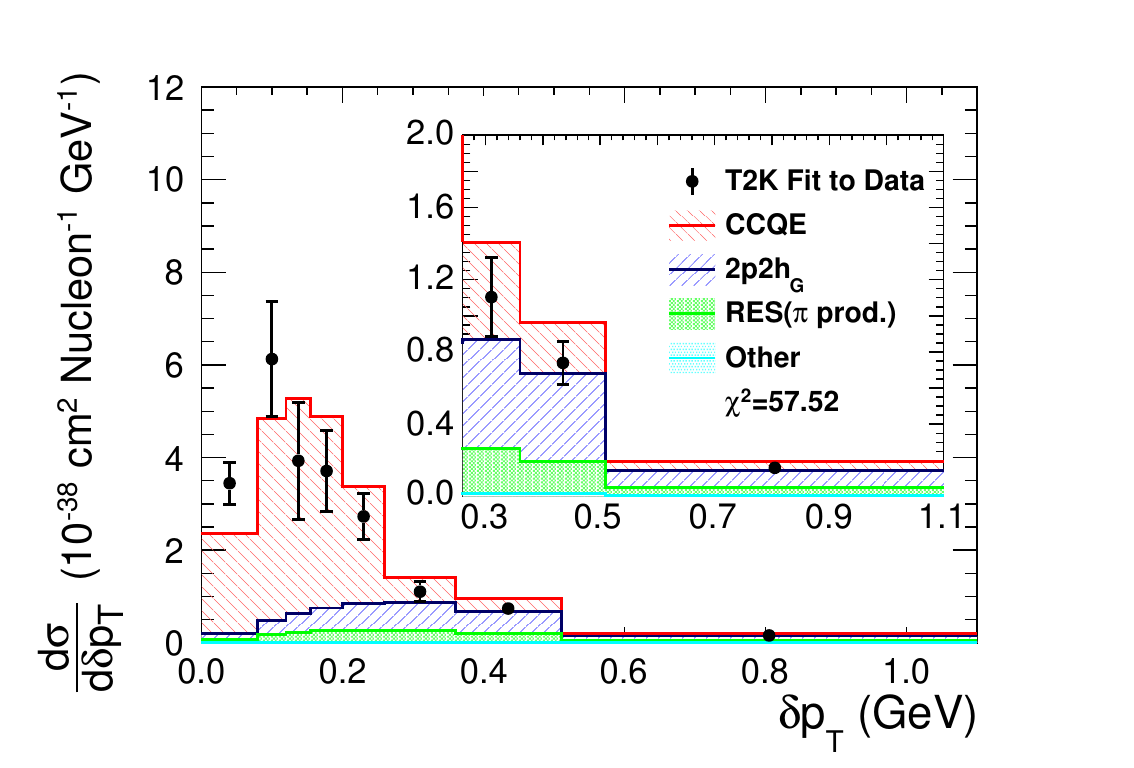}
\caption{$\delta p_{T}$ for $\mathcal{T} =1$} \label{fig:f}
\end{subfigure}
\vspace{-0.4cm}

\hspace{0.5cm}
\begin{subfigure}{0.52\textwidth}
\includegraphics[width=\linewidth]{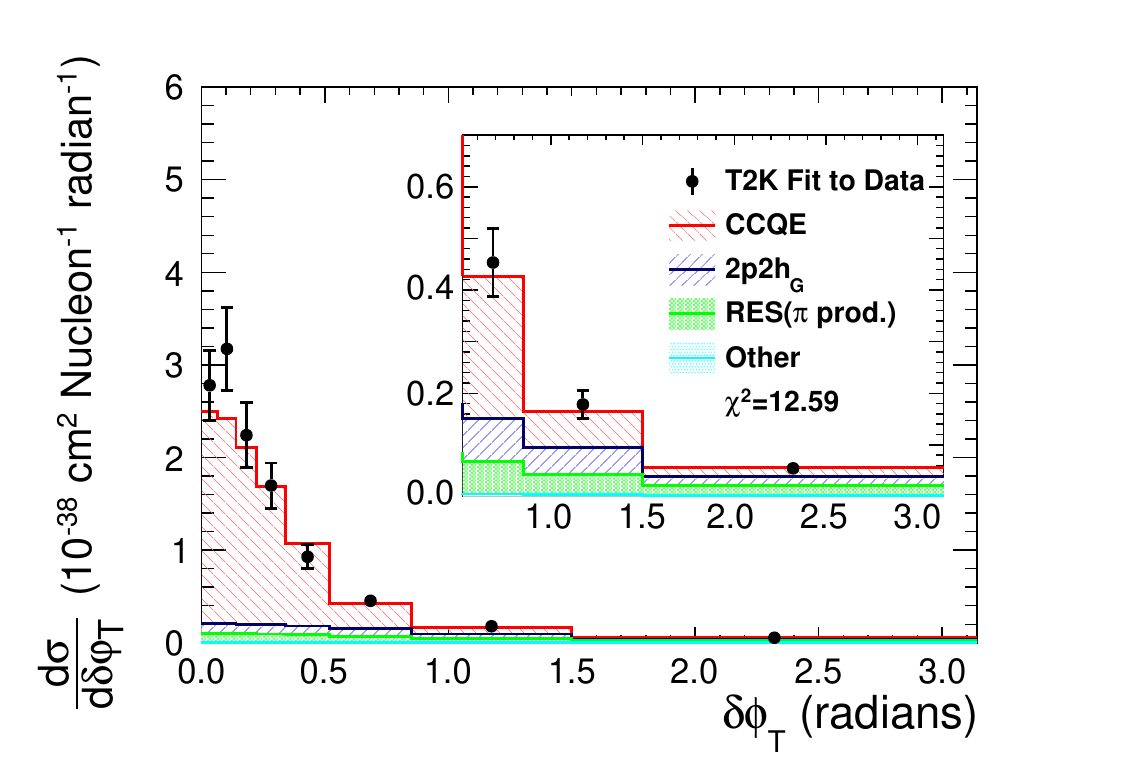}
\caption{$\delta\phi_{T}$ for $\mathcal{T} =0$} \label{fig:c}
\end{subfigure}\hspace{-1.1cm}
\begin{subfigure}{0.52\textwidth}
\includegraphics[width=\linewidth]{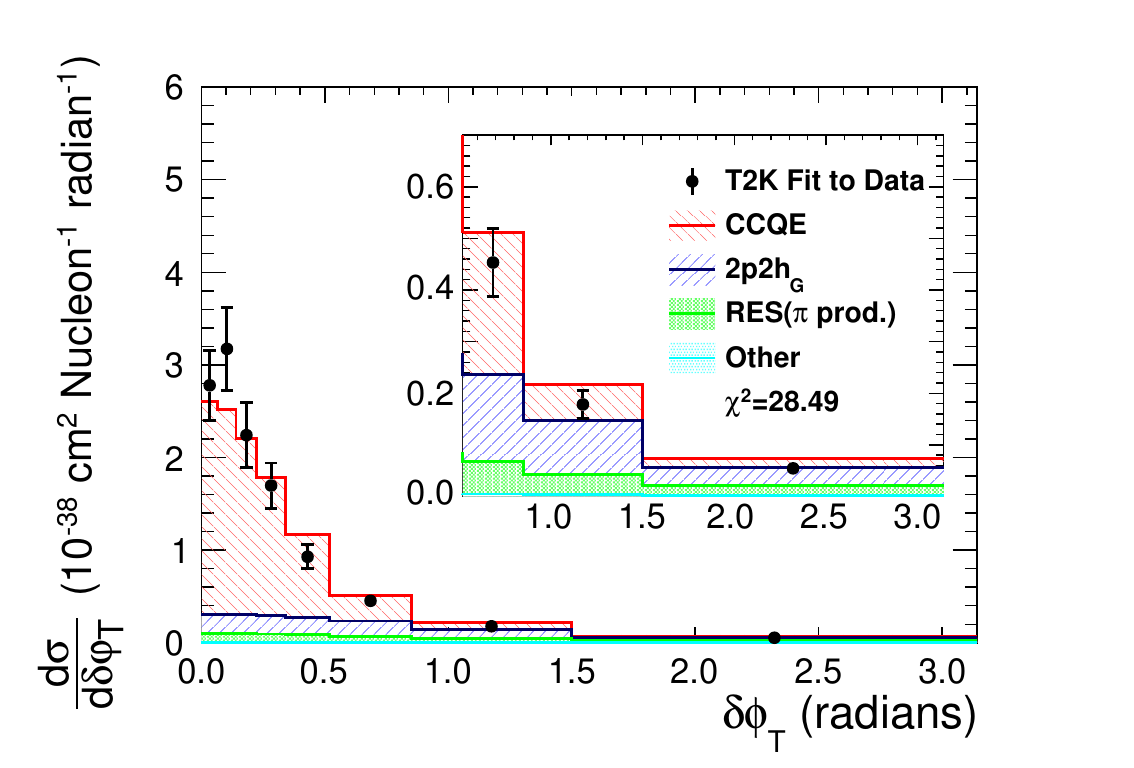}
\caption{$\delta\phi_{T}$ for $\mathcal{T} =1$} \label{fig:d}
\end{subfigure}
\vspace{-0.4cm}

\hspace{0.5cm}
\begin{subfigure}{0.52\textwidth}
\includegraphics[width=\linewidth]{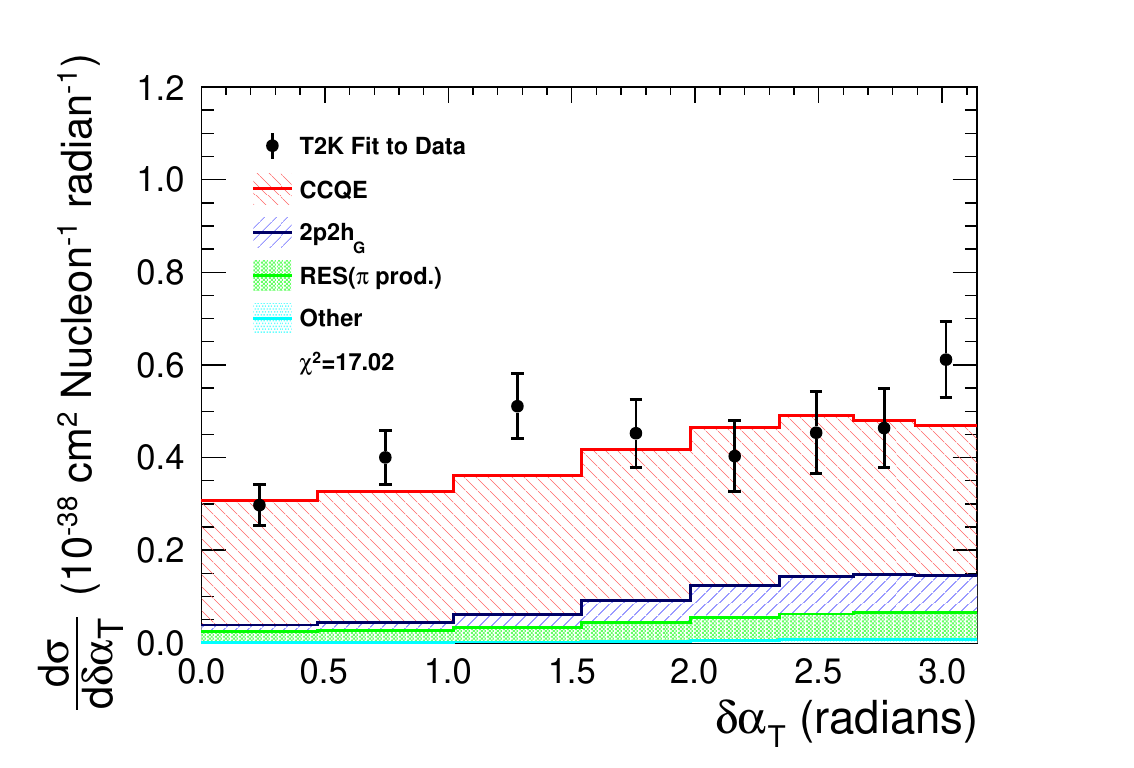}
\caption{$\delta\alpha_{T}$ for $\mathcal{T} =0$} \label{fig:a}
\end{subfigure}\hspace{-1.1cm}
\begin{subfigure}{0.52\textwidth}
\includegraphics[width=\linewidth]{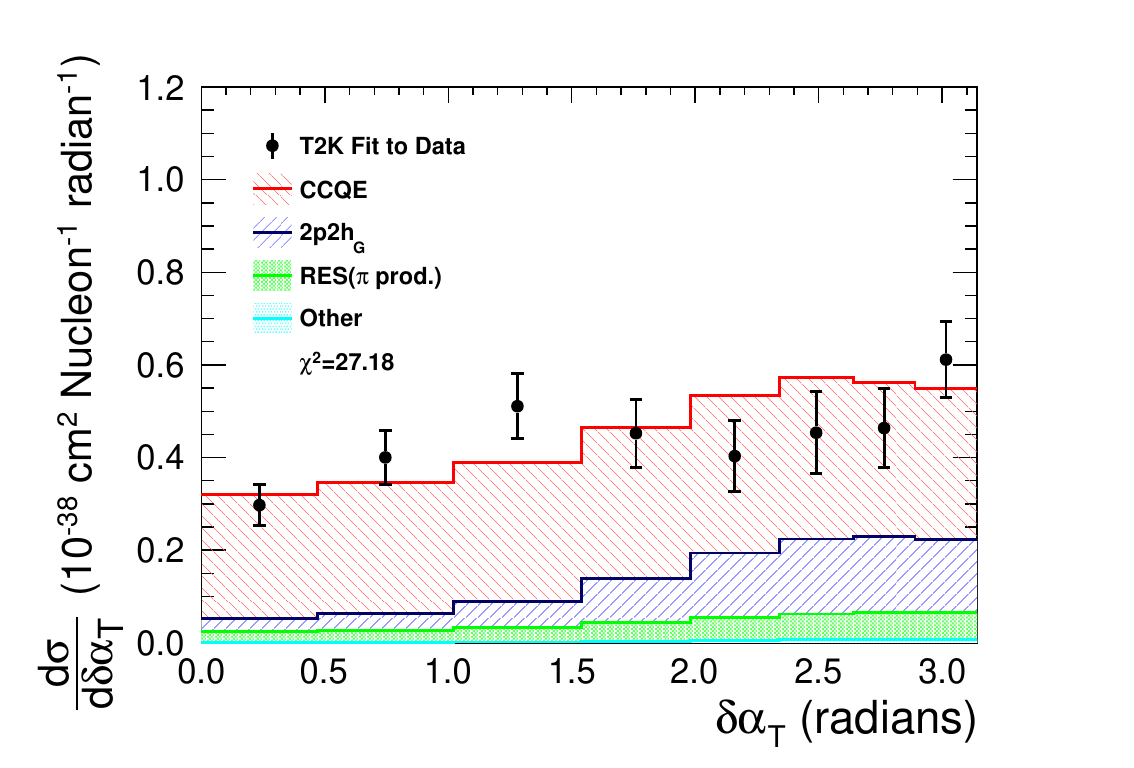}
\caption{$\delta\alpha_{T}$ for $\mathcal{T} =1$} \label{fig:b}
\end{subfigure}



\caption{The T2K measurement of the charged-current pionless cross-section on a C$_8$H$_8$ target as a function of the single-transverse variables within a restricted proton and muon kinematic phase-space (which is shown in Tab.~\ref{tab:phaseSpace}) \cite{Abe:2018pwo} compared with the results of GiBUU 2017 calculations. The plots on the left and right show the calculation with $\mathcal{T} =0$ and $\mathcal{T} =1$ respectively, each broken down by interaction mode. The inlays show a close-up of the tail regions of $\delta \phi_T$ and $\delta p_T$. The $\chi^2$ shown in the legends are calculated for eight degrees of freedom. } \label{fig:transverse}

\end{widepage}

\end{figure*}

Fig.~\ref{fig:transverse} shows the comparison of these calculated observables with the experimental ones, in the left column for $\mathcal{T}=0$ and on the right for $\mathcal{T}=1$. This comparison is made in the restricted proton and muon kinematic phase-space specified for the analysis, shown in Tab.~\ref{tab:phaseSpace}. 
\begin{table}[tb]
\centering
\begin{tabular}{ |c|c|c|c| } 
 \hline
$p_p$ & $\cos\theta_p$ & $p_\mu$ & $\cos\theta_\mu$  \\
 \hline
450-1000~MeV & $>0.4$ & $>250$~MeV & $>-0.6$ \\
\hline
\end{tabular}
\caption{The restricted proton and muon kinematic phase-space used in the analysis of transverse kinematic imbalance presented in Fig.~\ref{fig:transverse}. $p_p$, $\cos\theta_p$, $p_\mu$, $\cos\theta_\mu$ are the momentum and angle (with respect to the incoming neutrino) of the outgoing proton and muon respectively.}
 \label{tab:phaseSpace}
\end{table}

It is immediately seen that the $\mathcal{T}=1$ calculation disagrees with the result, particularly at large transverse kinematic imbalances (the tails of $\delta p_T$ and $\delta \phi_T$) where it becomes too large. This impression is borne out in the $\chi^2$ values, which are substantially lowered when the isospin configuration of the initial state is altered from $\mathcal{T}=1$ to $0$. Moreover, because the largest relative 2p2h contribution is highly concentrated in these large transverse imbalance regions, a similar improvement could not be obtained by alterations to the T2K flux normalization (the largest of the experimental systematic uncertainties). These measurements thus exhibit acute sensitivity to the different isospin assignments.

It should be noted that the shape of the model does not completely describe the result in the CCQE-dominated region of $\delta p_T$ (where $\delta p_T$ is not much larger than the Fermi momentum). However, this region is specifically sensitive to the initial state nucleon momentum distribution and alterations to this are not expected to strongly alter the tail of the distribution~\cite{Lu:2015tcr} so therefore have limited bearing in the determination of $\mathcal{T}$. 

It is possible that variations to the description of hadronic re-interactions inside the nucleus (`final state interactions', FSI) could also alter the high transverse imbalance regions. However, the description of FSI within GiBUU have been validated and tested across a wide class of reactions~\cite{Leitner:2009ke,Buss:2011mx} and are entirely consistent in their description of the physics of the processes involved. Moreover, it has been demonstrated (albeit for other models) that the limited proton kinematic phase-space employed by the analysis suppresses the impact of FSI alterations such that very large variations correspond to only small changes in the high transverse imbalance regions of the predicted distributions~\cite{Abe:2018uhf}. 

In these comparisons we essentially treat $\mathcal{T}$ as a free parameter. It is gratifying to see that indeed $\mathcal{T}=0$, the physical isospin of the groundstate of $^{12}$C, is favored.

\section{Summary and Conclusions}

A detailed understanding of 2p2h excitations in neutrino-nucleus interactions is essential to control the systematic uncertainties in long-baseline neutrino oscillation analyses. Previous measurements have struggled to give a consistent depiction of 2p2h, where the size of the contribution has often been degenerate with other effects (most markedly the neutrino flux prediction). Recent results from T2K, exploiting both muon and proton kinematics, allow an opportunity for a relatively transparent quantification of the 2p2h contribution.

GiBUU describes neutrino-nucleus 2p2h excitations by analogy to electron-nucleus scattering, which is much better measured, relating the neutrino structure functions to those for electrons by a simple factor where the only parameter to determine is the isospin, $\mathcal{T}$, of the initial state. Comparing to the latest T2K results, it is therefore satisfying to see that in the 2p2h-enhanced high transverse imbalance region, the result clearly favours $\mathcal{T}=0$ to $\mathcal{T}=1$, the former case showing excellent agreement with the results in the pertinent kinematic regions. Since the groundstate isospin of $^{12}$C is indeed $\mathcal{T}=0$, consistency of electron and neutrino scattering data is achieved. It will be interesting to look for this isospin-effect in experiments with $^{40}$Ar as a target, such as MicroBooNE and DUNE. Here $\mathcal{T} = 2$, so a stronger enhancement factor is expected.

\vspace{-1.5mm}
\section*{Acknowledgements}
\vspace{-1.5mm}
The authors would like to thank the members of the T2K Collaboration for producing and analyzing the results used in this manuscript. We are grateful to G.~Megias for helpful discussions regarding previous 2p2h model comparisons to data and to I.~Kakorin and V.~Naumov for identifying an error in the original $\chi^2$ reported. We acknowledge the support of CEA, CNRS/IN2P3 and P2IO, France; the Office of Science, Office of High Energy Physics, of the U.S. Department of Energy, under award Number DE-SC0015903; and the Alfred P. Sloan Foundation. We are grateful to the MSCA-RISE project JENNIFER, funded by EU grant n.644294, for supporting the EU-Japan researchers' mobility.

\makeatletter\onecolumngrid@push\makeatother

\setlength{\abovecaptionskip}{5pt}
\captionsetup[subfigure]{aboveskip=-6pt,belowskip=8pt}

\end{document}